# EXPONENTIAL PARALLELISM IN PRACTICE: A COMPARATIVE FEATURE ON QUANTUM COMPUTING AND INSTANTANEOUS NOISE-BASED LOGIC

LASZLO B. KISH

*Department of Electrical and Computer Engineering, Texas A&M University, TAMUS 3128, College Station, TX 77843, USA*

Exponential parallelism, a defining principle of advanced computational systems, holds promise for transformative impacts across several scientific and industrial domains. This feature paper provides a comparative overview of Quantum Computing (QC) and Instantaneous Noise-based Logic (INBL), focusing on their practical strengths, limitations, and applications rather than exhaustive technical depth. Both paradigms leverage exponentially large Hilbert spaces: quantum computing achieves this via quantum superposition, while INBL realizes it through the product space of classical noise processes. Quantum computers attain universality for all computational operations, whereas current INBL frameworks are universal only for Boolean logic; notably, essential superposition operations—such as AND and OR gates—are absent, precluding implementations of algorithms like Shor's. However, for certain problem classes where full universality is not required, INBL and quantum computing can offer equivalent time and hardware complexity, as observed with the Deutsch–Jozsa algorithm. Remarkably, for search tasks such as phonebook lookup, Grover's quantum algorithm provides a quadratic $O\left(\sqrt{N}\right)$ speedup compared to the classical approach, while INBL achieves an exponential speedup, requiring only logarithmic time in the size $N$ of the phonebook $O\left(\log N\right)$. Such INBL algorithms could, in principle, be adapted to quantum hardware to attain similar performance. Importantly, INBL hardware is considerably simpler, being implementable with modest modifications to conventional PC architectures equipped with a true random number generator, and it inherently avoids the decoherence and error correction challenges of quantum systems.

*Keywords:* Gate-based quantum computing; Instantaneous noise-based logic; status; comparison.

## 1. Introduction

This paper offers a bird's-eye overview comparing quantum computing (QC) [1-6] with its classical physical analogue, instantaneous noise-based logic.

## 2. On Gate-Based Quantum Computing [1]

QC promises to revolutionize computational capabilities for certain classes of problems, leveraging principles of quantum mechanics that have no analog in classical computation. The gate-based (or circuit-based) quantum computing model is the most widely adopted architecture for achieving this vision, offering theoretical universality, programmability, and potential exponential speedups for problems like factoring, molecular simulation, and search.

- *Qubits and Quantum States*

Central to all quantum computation is the qubit—the quantum equivalent of the classical bit—which exists as a vector in a two-dimensional Hilbert space. Unlike classical Boolean bits in a Turing machine, which can be only 0 or 1, qubits can be in superpositions, represented as $|\psi\rangle = \alpha|0\rangle + \beta|1\rangle$ where $\alpha$ and $\beta$ are complex amplitudes satisfying $|\alpha|^2 + |\beta|^2 = 1$. Collections of qubits form exponentially large joint state spaces, enabling rich phenomena such as entanglement and interference.

- *Quantum Gates and Circuits*

Gate-based quantum computers operate by applying a series of quantum logic gates—unitary transformations—on the qubits. These gates include single-qubit rotations and two-qubit entangling gates, the most important of which is the Controlled-NOT (CNOT) gate. Sequences of these gates form quantum circuits, which are analogous to classical logic circuits but operate under fundamentally different rules. Mathematically speaking, a universal set of gates can build, in principle, any arbitrary quantum algorithm; this universality is the foundation for theoretical models of quantum programming and exponential speedup when manipulating on exponentially large superpositions.

The principal challenge in quantum computing lies in formulating quantum algorithms that can operate not merely as abstract mathematical constructs but also as viable implementations on physical hardware.

- *Quantum universe: superposition of all binary numbers in the Hilbert space*

The concept of a "quantum universe" arises from the superposition principle applied within Hilbert space—a generalization of the classical bit string space into a multidimensional quantum domain. Each qubit inhabits a vector space spanned by the basis states $|0\rangle$ and $|1\rangle$; collectively, $n$ qubits occupy a $2^n$-dimensional Hilbert space, whose basis vectors correspond to all possible binary strings of length $n$.

Through the action of Hadamard gates on each qubit, a quantum computer initializes its register into an equal superposition of all binary numbers:

$$|\psi\rangle = \frac{1}{\sqrt{2^n}}\sum_{i=0}^{2^n-1} |x_i\rangle \tag{1}$$

where $|x_i\rangle$ represents an $n$ bit long string or binary number and the sum represents the superposition of all binary numbers from zero to $2^n$. This operation mathematically constructs a quantum state that simultaneously encodes all possible classical inputs, forming the foundation for exponential quantum parallelism. Mathematically, the quantum computing universe is represented by an exponential, $2^n$ x $2^n$, matrix thus it cannot be efficiently simulated by a classical Turing computer.

Physically, the quantum universe is not merely a statistical mixture; it is a coherent superposition supporting interference phenomena unique to quantum mechanics. Constructive and destructive interference patterns within this space, orchestrated through additional gate operations, allow quantum algorithms to amplify correct solutions while suppressing incorrect ones, effectively navigating the multitudes of possibilities that occupy the Hilbert space.

To further clarify the structure of the quantum universe, it is essential to note that the exponentially large joint Hilbert space of the quantum register arises from the tensor product of the individual qubit spaces. Each qubit is associated with a two-dimensional Hilbert space, and the complete quantum state space for *n* qubits is formally the tensor product

$$H = H_1 \otimes H_2 \otimes ...H_n \ , \tag{1}$$

yielding a $2^n$ dimensional space. This tensor structure underpins the unique ability of quantum systems to represent and manipulate the coherent superpositions of all $2^n$ possible binary strings simultaneously, as well as the foundation for entanglement and interference phenomena vital to quantum computation.

This mechanism is central to celebrated quantum algorithms such as Deutsch–Jozsa, Grover's search, and Shor's factoring, where quantum computation proceeds by exploring and manipulating the superposition of all binary numbers to yield computational advantages that surpass classical methods. Thus, the quantum universe— realized as a superposition over the $2^n$ binary numbers—stands as the stage for quantum computation in the Hilbert space.

It is important to emphasize that the Deutsch–Jozsa algorithm can also be realized [7] using Instantaneous Noise-based Logic [7-18], achieving computational complexity equivalent to that of its implementation on a quantum computer, see Section 3. Furthermore, the phonebook search algorithm based on INBL [8,9] demonstrates an exponential speedup compared to Grover's quantum search algorithm; notably, an analogous approach could be adapted for quantum computing, thereby attaining equivalent performance. However, it should be noted that, owing to the absence of certain gate functionalities, an INBL-based realization of Shor's algorithm remains unfeasible.

- *Quantum Parallelism and Interference*

Quantum circuits exploit "quantum parallelism": a superposition allows quantum gates to act simultaneously on all basis states, so a sequence of gates can process many possible input values in parallel. Properly designed, quantum algorithms choreograph interference such that correct outputs are amplified and incorrect ones are suppressed when qubit states are measured.

For example, Shor's algorithm for integer factoring and Grover's algorithm for search both use clever arrangements of quantum gate sequences to achieve speedups over the best-known classical algorithms. The art and science of quantum algorithm design lies in finding circuits that leverage these uniquely quantum phenomena.

- *Physical Realizations and Challenges*

Physical implementations of gate-based quantum computers range from superconducting circuits (IBM, Google), trapped ions, photonic chips, to spins in semiconductors. Each technology has tradeoffs in terms of scalability, fidelity, error rates, and coherence times—the ability of qubits to maintain quantum states before noise destroys computation. Decoherence and error correction remain key challenges.

- *Application & Outlook*

While there are some pessimistic views [2], it is indeed a fact that no practical "quantum advantage" applications for industry-critical problems are available yet. According to optimistic views (e.g. [3-6]), 2025 marks the start of the first generation of logical-qubit experiments that anticipate quantum advantage within the next 3–4 years. The financial industry, chemistry, and cryptography remain focus areas for early, commercial value. Industry confidence is high, with significant investments, but full-scale, fault-tolerant, gate-based quantum computers are still at least several years from deployment at industrial scale.

In summary, gate-based quantum computing in 2025 appears like it is defined by robust research momentum, hardware progress past 100 qubits, the practical debut and demonstration of logical qubits, and strong—but as yet unfulfilled—commercial ambitions for wide-scale utility in the late 2020s.

### 3. On Instantaneous Noise-based Logic

Instantaneous Noise-Based Logic (INBL) [7-18] is a classical computational paradigm designed to exploit the statistical properties of specific noise signals for efficient high-dimensional computation. The physical INBL encodes information using random telegraph waves enabling the creation of large product spaces that scale exponentially with the number of noise bits, equivalent to the complexity of quantum Hilbert spaces [7].

A noise-bit has two values, 0 (L) and 1 (H), each is represented by an independent, periodically clocked random telegraph wave (RTW), $R_0(t)$ and $R_1(t)$, respectively. In the simplest case they are random square waves with +1 and -1 values generated by a true random number generator, but alternative solutions also exist, such as different values for L and H [8], or complex values [9].

Strings and binary numbers in an INBL scheme with $n$ noise-bit resolution are products of the $n$ independent RTWs representing the corresponding bit values in the string. Thus strings and binary numbers are also RTWs [7].

- *Boolean Universality and Logic Gates*

INBL is provably universal for Boolean logic [10,11]: it supports the instantaneous implementation of core Boolean gates such as AND, OR, and NOT, establishing its ability to compute any Boolean function deterministically. These gates are realized via direct correlations within the product space of independent noise signals, without the requirement of time-averaging, leading to speed advantages over traditional noise-based schemes. Universality in Boolean logic guarantees that standard digital computation tasks can be modeled by INBL hardware. However the practical importance of this is currently negligible.

- *Superpositions and Gate Set Limitations*

A unique property of INBL is its ability to represent and process computational superpositions—distinct noise signals can be combined into higher-dimensional hyperspace vectors, similar to quantum superposition. Obviously, the numbers and strings that RTWs represent can be combined into superpositions, and algebraic operations on single noise-bits, strings, numbers or superpositions can offer parallel processing [7]. However, not all logic gates have been realized within the INBL superposition context: for binary systems, only NOT [7], CNOT [12,13], and XOR [14] gates are currently implemented during operations on superpositions. As an example, the simplest operation is a $NOT_k$ gate that acts on the $k$-th noise bit, $NOT_k = R_0(t)R_1(t)$ , a multiplication that works in the system with RTWs with +1 and -1 values.

AND and OR gates for general superpositions are still missing [13], a limitation that prevents INBL from supporting full universal computation in the superposed regime and excludes certain algorithms (such as Shor's factoring) that require those gates.

- *Extension to Ternary Systems*

Research in INBL has recently extended into ternary logic [15], where three distinct logic states are encoded by different noise signals. Though ternary INBL remains under active exploration, it promises richer expressiveness and the potential for more efficient parallel operations compared to binary noise-based logic.

- *Hardware and Practical Benefits*

RTWs, their products and superpositions make it easy to represent an INBL system in a binary classical computer with word length $n$ or $2n$ [8] depending on the types of RTWs used. This design circumvents issues fundamental to quantum devices—such as decoherence and error correction—by relying solely on robust classical physical processes. As a result, INBL hardware is simple, reliable, and more scalable using current technology compared to quantum computers. However, ideally a true random number generator is required to generate the RTWs to avoid unpleasant surprises at some algorithms. So far, such an issue has not yet been identified, but certainly exists.

- *The universe in INBL*

In Instantaneous Noise-Based Logic, the construction of the "universe" of all possible logic states—analogous to the quantum universe created by cascading Hadamard gates—is accomplished through the Achilles heel operation (e.g. [7,8]). This operation involves forming the superposition of all possible product strings of the reference noise signals (noise-bits) across the processor. Specifically, for $n$ noise-bits, the Achilles heel operation generates the complete sum of all $2^n$ product combinations, establishing a "universe state" which is a stochastic signal in the classical noise hyperspace. This state forms the foundational superposition from which deterministic parallel operations are possible and enables logic gates and memory addressing in INBL to act over the entire computational space at once, mirroring the effect of the Hadamard gate in quantum circuits.

It is important to note that the complete state space for $n$ noise-bits can also be described as a tensor product where each dimension corresponds to the stochastic process representing an individual noise-bit. Thus, the INBL hyperspace is realized as the tensor product

$$S = S_1 \otimes S_2 \otimes ... S_n \ ,\tag{2}$$

encompassing all the $2^n$ possible composite noise states. This explicit product structure enables deterministic parallel operations over the entire computational space, mirroring the quantum case but realized with classical variables

This deterministic creation of the universal superposition enables INBL to implement parallel algorithms in a manner analogous to Hadamard-based universes in quantum computing, but using robust classical noise processes and simple hardware without the need for quantum coherence.

- *Applications, demonstrations*

INBL has been experimentally validated through prototype demonstrations that showcase its capability for deterministic, high-speed computation across exponentially large state spaces using classical noise signals. These demonstrations establish INBL as a practical classical analog to quantum parallelism and achieving comparable effects without invoking quantum coherence.

Notable applications include:

*i) Exponential phonebook search*: INBL has been used [16] to carry out deterministic dataset lookups across exponentially (vs the number of noise bits) large, unsorted search spaces. This "phonebook" problem, where a specific entry must be found among all possible combinations, was solved in polynomial time vs the number of noise-bits, due to the intrinsic parallelism of noise-vector operations. This result demonstrates a classical analog of Grover-like parallel search without probabilistic measurement. It utilizes classical entanglement for exponential speedup. In principle, quantum algorithms should also be able to use this scheme; the question is the relevant hardware.

*ii) Deterministic realization of the Deutsch–Jozsa algorithm*: INBL implementations [7] have replicated the structure of the Deutsch–Jozsa quantum algorithm, using correlations among noise vectors to evaluate global Boolean properties in polynomial time. Unlike its quantum counterpart, the INBL solution operates deterministically with immediate logic outcomes and robust signal-to-noise immunity.

*iii) "Hat problem" and exponential configuration search*: In these tests, INBL successfully executed parallel operations across a full noise hyperspace to remove or identify specific integer configurations—such as finding an element "drawn from a hat."

The demonstration confirmed that INBL can execute computation over all $2^n$ possible binary numbers instantaneously, analogous to operating on a superposed quantum state [8,17].

These results support the contention that INBL can emulate some of the exponential complexity advantages of quantum computation while preserving classical determinism and hardware simplicity.

- *Note on the "compressions" in Instantaneous Noise-based logic*

From a different angle, we can realize that there are two different but essential compressions in the INBL system. These aspects are also behind the speedup demonstrated by computer simulations [8,17].

i) The universe has the sum of $2^n$ RTWs as the numbers represented by INBL are RTWs themselves. That amounts to an exponentially large amplitude which an analog computer would not be able to process. However digital computers operate bits which yield a logarithmic transformation versus but significance thus INBL by digital computers is feasible.

ii) Bandwidth compression. Theoretically, the INBL scheme could be represented by sinusoidal waves instead of RTWs [18]. However, to avoid losing information, the frequencies of the sinusoidals must be chosen by an exponential scale which would cause an exponential time complexity (the ratio of the highest and the lowest frequencies in the system). On the other hand, the RTWs have the same clock frequency. And multiplying them does not increase this frequency. The price is that it is as difficult to analyze an $O\left(2^n\right)$ large superposition in INBL as in QC except for the statistical measurement output in quantum which is a further complication.

## 4. Quick Comparison of Quantum Computing and Instantaneous Noise-Based Logic

While both QC and INBL operate on exponentially scalable state spaces, their physical foundational mechanisms differ profoundly. QC leverages quantum superposition and entanglement, realized through delicate quantum states in complex physical systems, whereas INBL exploits classical noise superpositions built from independent stochastic signals, achieving parallelism deterministically without invoking quantum phenomena. The universe in both QC and INBL are achieved by tensor products which indicates the deep structural analogy between quantum and INBL universes, emphasizing their foundation as tensor product spaces.

In essence, INBL can reproduce certain exponential features of quantum computing deterministically using classical systems, including superpositions, parallel operations and (classical) entanglement, but without the universality that grants QC its broader computational potential. Conversely, QC's probabilistic nature and physical fragility continue to pose engineering challenges, while INBL provides a physically robust classical architecture for a subset of parallel computational problems, see Table 1.

Unlike quantum hardware, which must contend with precise control of quantum states and complex error correction, INBL solutions run on ordinary computers with long word length (several thousands bits) equipped with a true random number generator. The long word length is achievable also by special software as speed is unimportant due to the exponential speedup by INBL.

Table 1. Quick comparison of the two schemes.

| Aspect | Quantum Computing (QC) | Instantaneous Noise-Based Logic (INBL) |
|---|---|---|
| **Physical Basis** | Quantum superposition and entanglement in qubits | Classical superpositions of the products of independent noise processes and classical entanglement |
| **State Space** | $2^n$-dimensional Hilbert space of qubit states | $2^n$-dimensional space of noise-bit product-strings |
| **Universality** | Fully universal for quantum gates (unitary operations) | Universal only for Boolean logic; lacks AND/OR operations for superpositions. Currently it has only NOT, XOR, and CNOT |
| **Speed Characteristics** | Probabilistic, governed by interference and measurement post-selection | Deterministic, operates via direct algebraic manipulation of noise bits, their strings and the superpositions of strings. |
| **Error and Decoherence** | Subject to decoherence, requiring extensive error correction | Immune to decoherence; inherently stable due to classical noise basis. |
| **Hardware Complexity** | Requires cryogenic environments, quantum gates, and precise isolation | Implementable with modified conventional digital or analog circuits |
| **Scalability** | Limited by quantum error correction overheads | Readily scalable with classical hardware components |
| **Algorithmic Examples** | Shor's, Grover's, and quantum simulation algorithms | Deterministic analogues of Deutsch–Jozsa and exponential "hat" or search problems |
| **Performance Regime** | Probabilistic exponential advantage for specific problem classes | Deterministic polynomial or exponential advantage for non-universal problems |
| **Maturity and Outlook** | Experimental, with logical qubits emerging post-2025 | Computer simulations exist. Prototype demonstrations achievable with existing hardware technology. Requires true random number generator and long word length that is achievable also by special software. |